\documentclass[journal]{IEEEtran}

\usepackage[T1]{fontenc}
\usepackage[latin9]{inputenc}
\usepackage{graphicx}
\usepackage{bm}
\usepackage{textcomp}
\usepackage[english]{babel}
\usepackage[noadjust]{cite}
\usepackage{notes2bib}
\bibnotesetup{
note-name = ,
use-sort-key = false
}


\usepackage{fancyhdr}
\usepackage[colorlinks,citecolor=black,linkcolor=black,urlcolor=blue]{hyperref}

\begin{document}

\title{Visible-spanning flat supercontinuum for astronomical applications}

\author{Aakash~Ravi, Matthias~Beck, David~F.~Phillips, Albrecht~Bartels, Dimitar~Sasselov, Andrew~Szentgyorgyi, and Ronald~L.~Walsworth
\thanks{A. Ravi is with the Department
of Physics, Harvard University, Cambridge, MA 02138, USA e-mail: aravi@physics.harvard.edu}
\thanks{M. Beck and A. Bartels are with Laser Quantum GmbH, Max-Stromeyer-Str. 116, 78467 Konstanz, Germany.}
\thanks{D. F. Phillips, D. Sasselov and A. Szentgyorgyi are with the Harvard-Smithsonian Center for Astrophysics, 60 Garden St., Cambridge, MA 02138, USA.}
\thanks{R. L. Walsworth is with the Department
of Physics, Harvard University, Cambridge, MA 02138 and also with the Harvard-Smithsonian Center for Astrophysics, 60 Garden St., Cambridge, MA 02138, USA.}
\thanks{(Dated: \today)}
}

\markboth{Journal of Lightwave Technology}
{Shell \MakeLowercase{\textit{et al.}}: Bare Demo of IEEEtran.cls for IEEE Journals}

\maketitle

\begin{abstract}
We demonstrate a broad, flat, visible supercontinuum spectrum that is generated by a
dispersion-engineered tapered photonic crystal fiber pumped by a 1 GHz repetition rate turn-key Ti:sapphire laser outputting $\sim$~30~fs pulses at 800
nm. At a pulse energy of 100 pJ, we obtain an output spectrum
,that is flat to within 3 dB over the range 490-690 nm with a blue
tail extending below 450 nm. The mode-locked laser combined with
the photonic crystal fiber forms a simple visible frequency comb system
that is extremely well-suited to the precise calibration of astrophysical
spectrographs, among other applications.
\end{abstract}


\thispagestyle{fancy}

\section{Introduction}

\IEEEPARstart{D}{oppler} spectroscopy of stars using high-resolution astrophysical
spectrographs enables the detection of
exoplanets through measurement of periodic variations in the radial velocity of the host star \cite{Fischer2016}. Since such measurements are inherently
photon-flux-limited and require spectral sensitivity much better than the resolution
of the spectrograph, they require combining information from thousands
of just-resolved spectral lines across the passband of the instrument. Doing this
reliably over orbital timescales requires an extremely stable calibration
source with a large bandwidth and uniform spectral coverage. State-of-the-art
spectrographs used in Doppler exoplanet searches such as HARPS \cite{Mayor2003}
and HARPS-N \cite{Cosentino2012} operate in the visible wavelength
range (400-700 nm). Laser frequency combs are well suited to calibrating
these instruments, and several ``astro-comb'' designs have been successfully
demonstrated to date (see Ref. \cite{McCracken2017a} and references
therein). 

Existing astro-comb architectures are based on near-infrared source combs (e.g.
Ti:sapphire, Yb/Er fiber), so providing visible calibration light
for an astrophysical spectrograph typically requires a nonlinear optical
element to coherently shift and broaden the source comb radiation. Early astro-combs derived from Ti:sapphire source combs relied
on second harmonic generation \cite{Benedick2010,Phillips2012}
but had limited utility due to extremely low output bandwidth ($\sim$
15 nm). This is a serious shortfall because the exoplanet detection
sensitivity depends on the bandwidth of the observed stellar light that is calibrated. A calibration source with constant line spacing over a larger bandwidth can therefore enable more precise determination of stellar Doppler shifts.

A better alternative to frequency doubling is
to pump a highly nonlinear photonic crystal fiber (PCF) with the source
comb to take advantage of supercontinuum generation \cite{Dudley2006},
an effect where a narrowband high-intensity pulse experiences extreme
spectral broadening as a result of interactions with the medium through
which it propagates. The dispersion and nonlinearity of PCFs may be
engineered via suitable changes in geometry; for example, two
commonly used parameters are the pitch and the diameter of the air
holes. PCFs also typically exhibit very high nonlinearities compared
to standard optical fibers due to their small effective mode field
diameters. As a result of these attractive properties, broadband supercontinuum
generation using PCFs has found many applications,
from optical coherence tomography \cite{Hartl2001} to carrier-envelope phase
stabilization of femtosecond lasers \cite{Jones2000}, as well as calibration of 
astrophysical spectrographs \cite{Wilken2012,Probst2013,McCracken2017,Glenday2015,Ravi2017}.

Beyond spectral bandwidth, another property that is desired for astronomical
calibration applications is spectral flatness, i.e., low intensity
variation across the band of the calibration source. Spectral flatness
is valued because calibration precision is both shot-noise- and CCD-saturation-limited. Therefore the largest photoelectron count
below CCD saturation per exposure provides the optimal calibration. This condition is best fulfilled for a uniform intensity distribution over all the comb peaks \cite{Probst2013}.

In the present work, we show that careful dispersion-engineering
of a PCF by tapering enables the production of a very
broad, flat, and high-intensity optical supercontinuum spectrum from pump radiation emitted
by a turn-key Ti:sapphire comb. Previous attempts to produce such
a supercontinuum from a Ti:sapphire comb have reported relatively low spectral
coverage (500-620 nm) and large intensity variations ($\sim$ 10 dB)
across the band \cite{Glenday2015}. A flat 235 nm-wide visible supercontinuum
has been demonstrated using a Yb:fiber comb \cite{Probst2014}, but at
the expense of losses incurred by spectral flattening using a spatial light modulator \cite{Probst2013}.

\section{Fiber Geometry and Parameters}

\begin{figure}
\includegraphics[width=1\columnwidth]{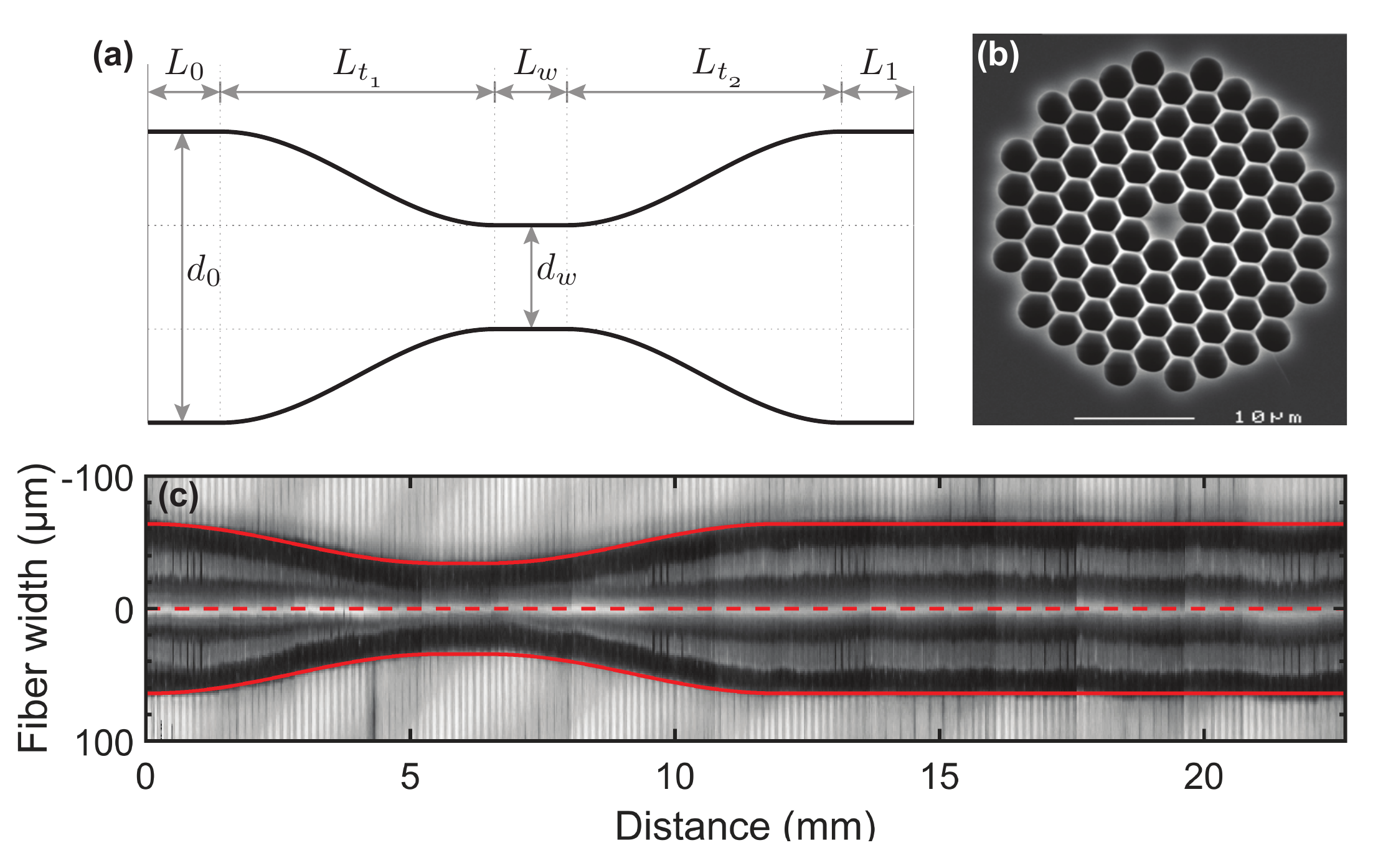}

\caption{\label{fig:geometry}(a) General tapered photonic crystal fiber (PCF) geometry showing varying core size vs. length. The quantities $d_0$, $d_w$, $L_{0,1}$, $L_{t_{1,2}}$, and $L_w$ parametrize the geometry; see text for details. (b) SEM micrograph (courtesy of NKT Photonics
A/S) of end face of NL-2.8-850-02 PCF. (c) Composite microscope
image of fabricated tapered PCF. Overlaid red curves represent the
proposed geometry, scaled by
the cladding/core size ratio.}
\end{figure}

In our application, we consider a PCF with a 2.8 \textmu{}m
core diameter and an 850 nm zero dispersion wavelength (NKT Photonics
NL-2.8-850-02). The cross section of the fiber is shown in Figure
\ref{fig:geometry}b. We chose a large core diameter to facilitate
coupling and have the zero dispersion wavelength (ZDW) near our 800
nm pump wavelength. Changing the PCF core diameter as a function of distance along the fiber 
via tapering modifies both the dispersion and nonlinearity of the
PCF. Sometimes termed ``dispersion
micromanagement'' in the literature, such techniques have been pursued before with
PCFs to enable generation of light with increased
bandwidth and flatness \cite{Lu2005,Lu2006,Kudlinski2006,Pal2007}, but not targeted
specifically toward uniform visible wavelength coverage using GHz repetition rate lasers.

To design a device capable of producing a flat visible-spanning spectrum when pumped with 800 nm femtosecond pulses, we model optical pulse propagation in the PCF by solving the generalized nonlinear Schr\"{o}dinger equation (GNLSE) as outlined in Appendix A. In our design, we consider a specific taper geometry, as shown in
Figure \ref{fig:geometry}a. The core diameter $d$ changes smoothly over
the tapers with a cosine function $d\left(\zeta\right)=d_{i}+\frac{1}{2}\left(1-\cos\pi\zeta\right)\left(d_{f}-d_{i}\right)$
over $\zeta\in\left[0,1\right],$ where $d_{i}$ and $d_{f}$ are,
respectively, the initial and final core diameters of the down-
or up-taper. More explicitly, $d_i=d_0$, $d_f=d_w$ for the down-taper and $d_i=d_w$, $d_f=d_0$ for the up-taper. The nondimensionalized variable $\zeta$ parametrizes
the distance along the taper, i.e., $\zeta=\left(z-L_{0}\right)/L_{t_{1}}$ for the down-taper and $\zeta=\left(z-\left(L_{0}+L_{t_{1}}+L_{w}\right)\right)/L_{t_{2}}$ for the up-taper.

\begin{figure}
\includegraphics[width=1\columnwidth]{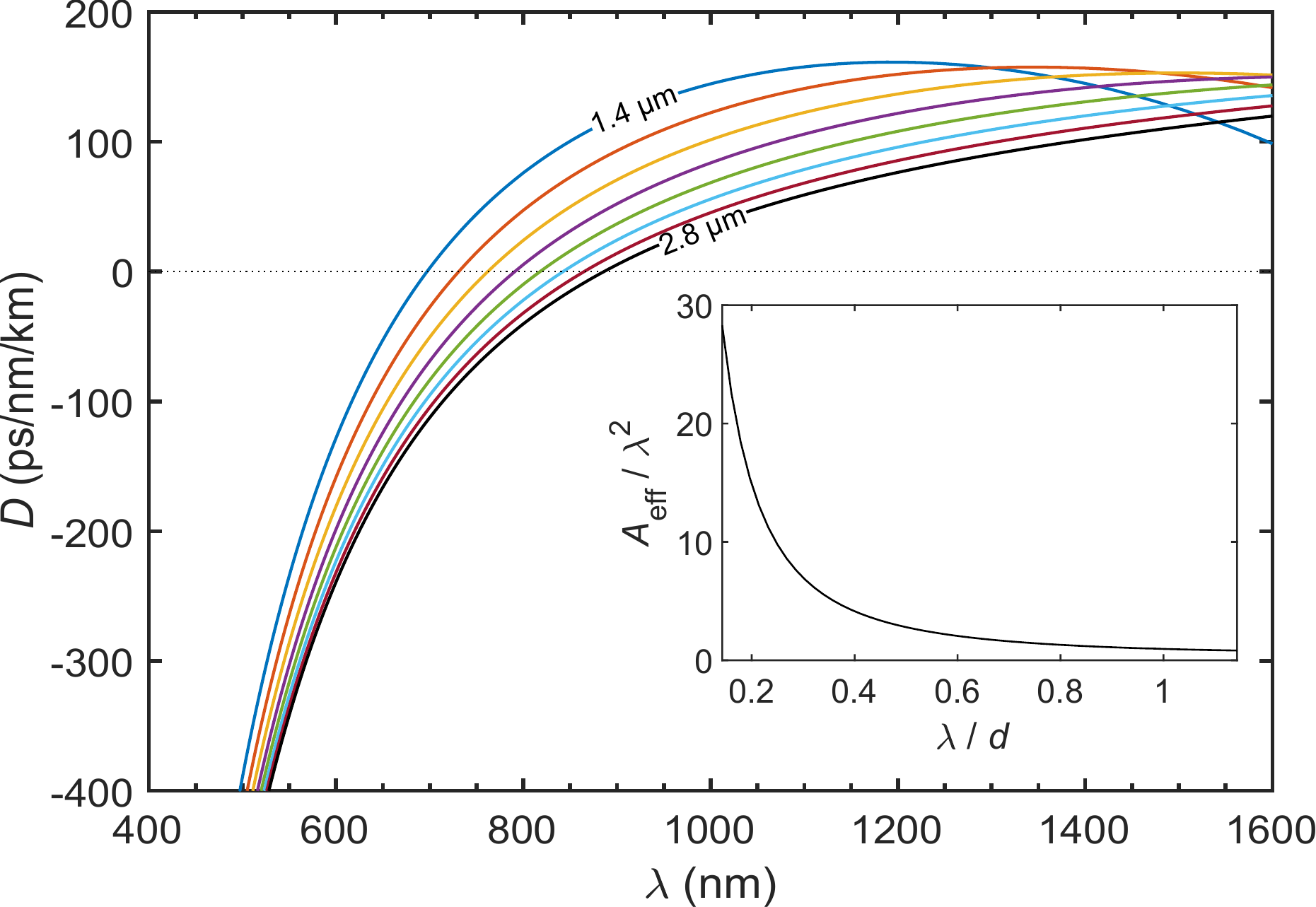}
\caption{\label{fig:fiber_properties}Simulated PCF properties. Main plot:
chromatic dispersion curves for varying core diameter $d$ (curves
spaced apart by $\Delta d=\textrm{0.2 \textmu{}m}$). Inset: normalized effective area vs. normalized wavelength.}
\end{figure}

Over the desired range of core diameters ($\textrm{1.4 \textmu{}m}\leq d\leq\textrm{2.8 \textmu{}m}$),
we compute the PCF dispersion and
nonlinearity as inputs for the pulse propagation calculations. We use a commercial finite-difference mode solver
(Lumerical MODE Solutions) to calculate the PCF properties. Figure
\ref{fig:fiber_properties} shows the chromatic dispersion $D=-\frac{\omega^{2}}{2\pi c}\frac{\partial^{2}\beta}{\partial\omega^{2}}$
versus wavelength $\lambda$ as a function of core diameter $d$ ($\beta$ is the propagation constant of the fundamental mode and $\omega$ is the angular frequency of light). 
Additionally, the inset of Figure \ref{fig:fiber_properties} shows
the modal area $A_{\mathrm{eff}}$ calculation for the PCF. This calculation
is required only for a single core diameter because $A_{\mathrm{eff}}$
respects the scale invariance of Maxwell's equations, as pointed out
in Ref. \cite{Laegsgaard2012}, whereas dispersion does not. In subsequent
results where we solve the GNLSE in a tapered PCF, the dispersion was interpolated to the local fiber
diameter from a pre-computed set of dispersion curves spaced apart
by $\Delta d=\textrm{0.05 \textmu{}m}$. In addition, the effective
area was evaluated for each PCF diameter according to the computed
curve in terms of the normalized variables shown in the inset of Figure
\ref{fig:fiber_properties}.

\section{\label{sec:Proposed-Design}Tapered Fiber Design}

\begin{figure*}
\includegraphics[width=2\columnwidth]{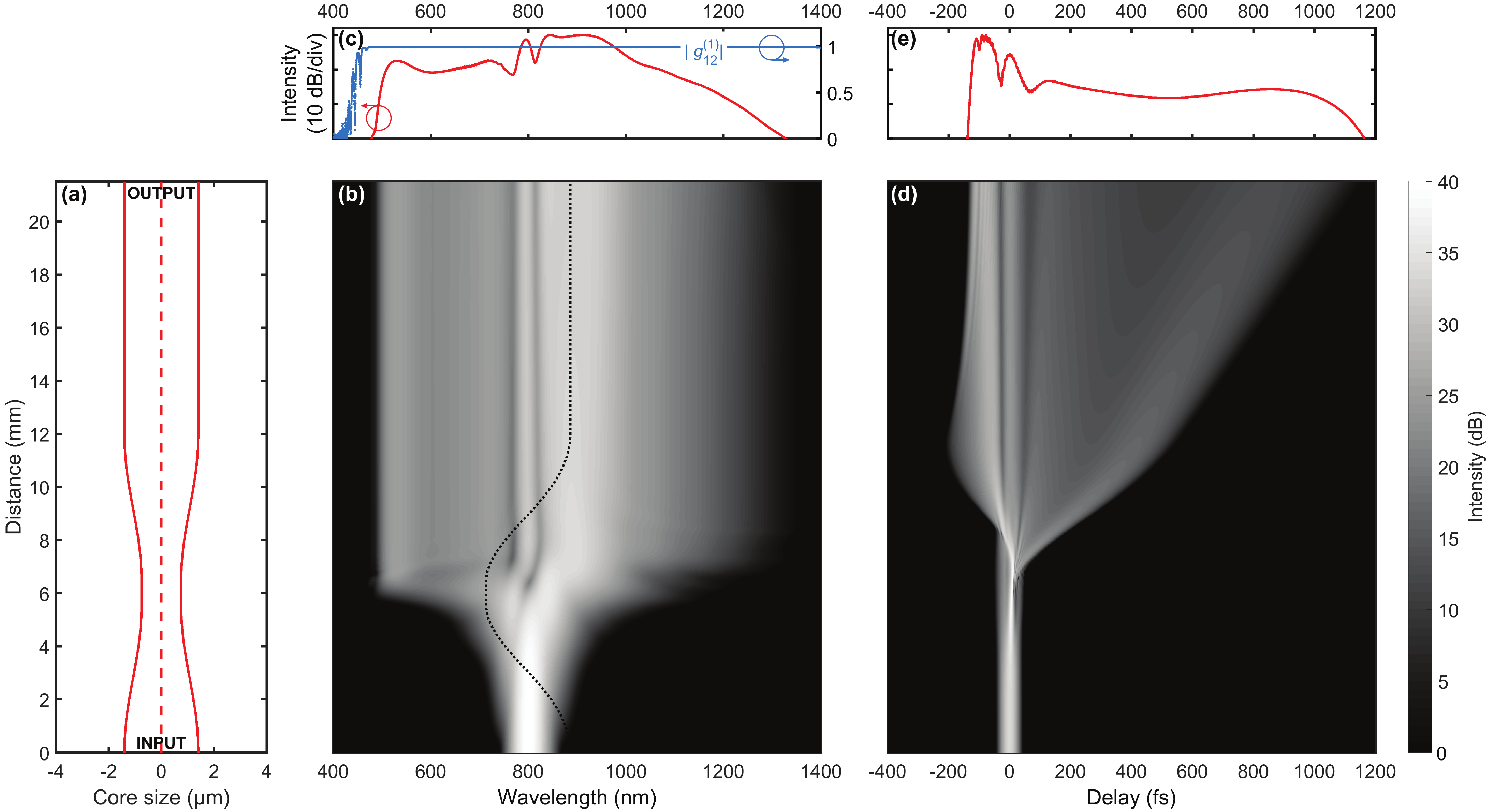}

\caption{\label{fig:mainsim}Simulated spectral and temporal evolution in a tapered PCF of a 215 pJ, 27 fs Gaussian pulse at 800 nm. (a) PCF core size
vs. longitudinal distance (0 is the fiber input face). (b) Pulse evolution~vs.~distance
in the spectral (wavelength) domain. The PCF's computed zero dispersion wavelength (ZDW) is superimposed
as a dotted black curve. (c) Pulse spectrum and magnitude of the first-order
coherence $g_{12}^{\left(1\right)}$ at the fiber output face (d) Pulse
evolution vs. distance in the time domain. (e) Pulse time trace at
fiber output face.}
\end{figure*}
We design a taper geometry for our PCF with the following parameters: $d_0=\textrm{2.8 \textmu{}m}$,
$d_w=\textrm{1.5 \textmu{}m}$, $L_0=\textrm{50 \textmu{}m}$, $L_{t_{1}}=L_{t_{2}}=\textrm{5.5 mm}$,
$L_w=\textrm{900 \textmu{}m}$ and $L_{1}=\textrm{9.55 mm}$, as shown
in Figure \ref{fig:mainsim}a. These geometric parameters were manually chosen to produce a supercontinuum with low intensity variation in the visible wavelength range while trying to push the blue edge of the supercontinuum to wavelengths as short as possible. A design derived using an optimization method with a merit function capturing these desired qualities may lead to enhanced performance, but is outside the scope of this work.

The pump radiation for the PCF is sourced from a taccor comb (Laser Quantum). This system is a turn-key 1 GHz Ti:sapphire
mode-locked laser with repetition rate and carrier-envelope offset
stabilization. The Ti:sapphire mode-locked laser is based on the technology described in Ref. \cite{Ma2004}, and is referenced to a GPS-disciplined Rb frequency standard, yielding a fractional stability of better than 10$^{-\textrm{\footnotesize 12}}$ \cite{Li2008}. Therefore, the source comb will not limit the frequency stability of the astro-comb \cite{Ravi2017}. The comb center wavelength is $\sim$ 800 nm and the
laser output bandwidth is approximately 35 nm, so the transform-limited
intensity full width at half maximum (FWHM) is 27 fs for a Gaussian
pulse shape.

By integrating the GNLSE (see Appendix A) with a fourth-order Runge-Kutta method, we investigate idealized pulse evolution in the tapered PCF assuming
an initial Gaussian pulse shape of the form $\sqrt{P_{0}}\exp\left(-2\ln2\left(T/T_{0}\right)^2\right)$.
Here, $T_{0}$ is the pulse FWHM, $P_{0}\approx\textrm{0.94}E_{p}/T_{0}$
is the pulse peak power, and $E_{p}$ is the pulse energy. Figure \ref{fig:mainsim}b
and d show the pulse evolution in the spectral and time domain, respectively
for a 215 pJ pulse injected into the tapered PCF, whose geometry is represented in Figure \ref{fig:mainsim}a. Panels c and e show the
spectrum and time trace of the pulse at the fiber output. Panel c
also shows the magnitude of the first-order coherence $g_{12}^{\left(1\right)}$
of the output (see Appendix B for more details). Focusing on the panels
b and d, we see that the initial dynamics correspond to symmetric
spectral broadening and an unchanged temporal envelope associated
with self-phase modulation  \cite{Dudley2010}. This is because the pump pulse is initially propagating in the
normal dispersion regime (i.e., chromatic dispersion $D<\textrm{0}$). As the PCF narrows, the pulse crosses over into the anomalous ($D>\textrm{0}$) dispersion regime. Here, perturbations
such as higher-order dispersion and Raman scattering
induce pulse break-up in a process called soliton fission \cite{Dudley2010}. Subsequent to fission, the solitons transfer some of their energy to dispersive waves (DW) propagating
in the normal dispersion regime according to a phase matching condition
\cite{Akhmediev1995}. DW generation (also known as fiber-optic Cherenkov radiation), the phenomenon responsible for generating the short-wavelength radiation in this case, has been thoroughly studied in the past  \cite{Chang2010a,Chang2011}. The supercontinuum bandwidth is dependent on
the pump detuning from the ZDW, so typically one can achieve larger
blue shifts in constant-diameter PCFs by using smaller core sizes, but at the expense of a spectral gap opening up between the pump and DW radiation
\cite{Dudley2010}. This issue can be addressed by using a tapered geometry, where the sliding phase matching condition along the narrowing PCF can generate successively shorter wavelength DW components \cite{Lu2005,Lu2006,Kudlinski2006,Pal2007}. Beyond the taper waist, the spectrum stabilizes as a result of the relaxation of the light confinement and corresponding reduction in nonlinearity.
This enables us to obtain a structure-free, flat band of light spanning
500-700 nm (Fig.~\ref{fig:mainsim}c). The light is also
extremely coherent ($\left|g_{12}^{\left(1\right)}\right|\approx\textrm{1}$)
across the whole spectral region containing significant optical power.

\section{Experimental Results}

Following the above design, the tapered PCF was fabricated from NL-2.8-850-02 fiber
by M.\ Harju at Vytran LLC on a GPX-3000 series optical fiber glass
processor using a heat-and-pull technique. Measurements from a micrograph of the
fabricated device (Fig.~\ref{fig:geometry}c)
agree well with the design geometry. We tested the tapered PCF by
clamping it in a V-groove mount and pumping it with light from a taccor
comb (Laser Quantum), recording the output spectrum on an optical
spectrum analyzer (OSA). An 8 mm effective focal length aspheric lens
(Thorlabs C240TME-B) was used for in-coupling, and a microscope objective
(20$\times$ Olympus Plan Achromat Objective, 0.4 NA, 1.2 mm WD) was
used for out-coupling. A dispersion compensation stage based on chirped
mirrors was used prior to coupling to counteract the chirp induced
by downstream optical elements and bring the pulse FWHM at the fiber
input close to its transform limit. A half-wave plate was also inserted
into the beam path before the in-coupling lens for input polarization
control. Coupled power was measured with a thermal power meter just
after the out-coupling microscope objective. Light was coupled into
the OSA via multimode fiber (Thorlabs FG050LGA) using a fixed focus
collimator (Thorlabs). We recorded spectra for a variety of femtosecond 
laser pump powers, optimizing the coupling before 
each measurement. For this series of measurements, the wave plate angle $\phi_{\lambda/2}$ was kept
constant. All measurements were performed with the same
equipment, the same dispersion compensation, and on the same day.

\begin{figure}
\includegraphics[width=1\columnwidth]{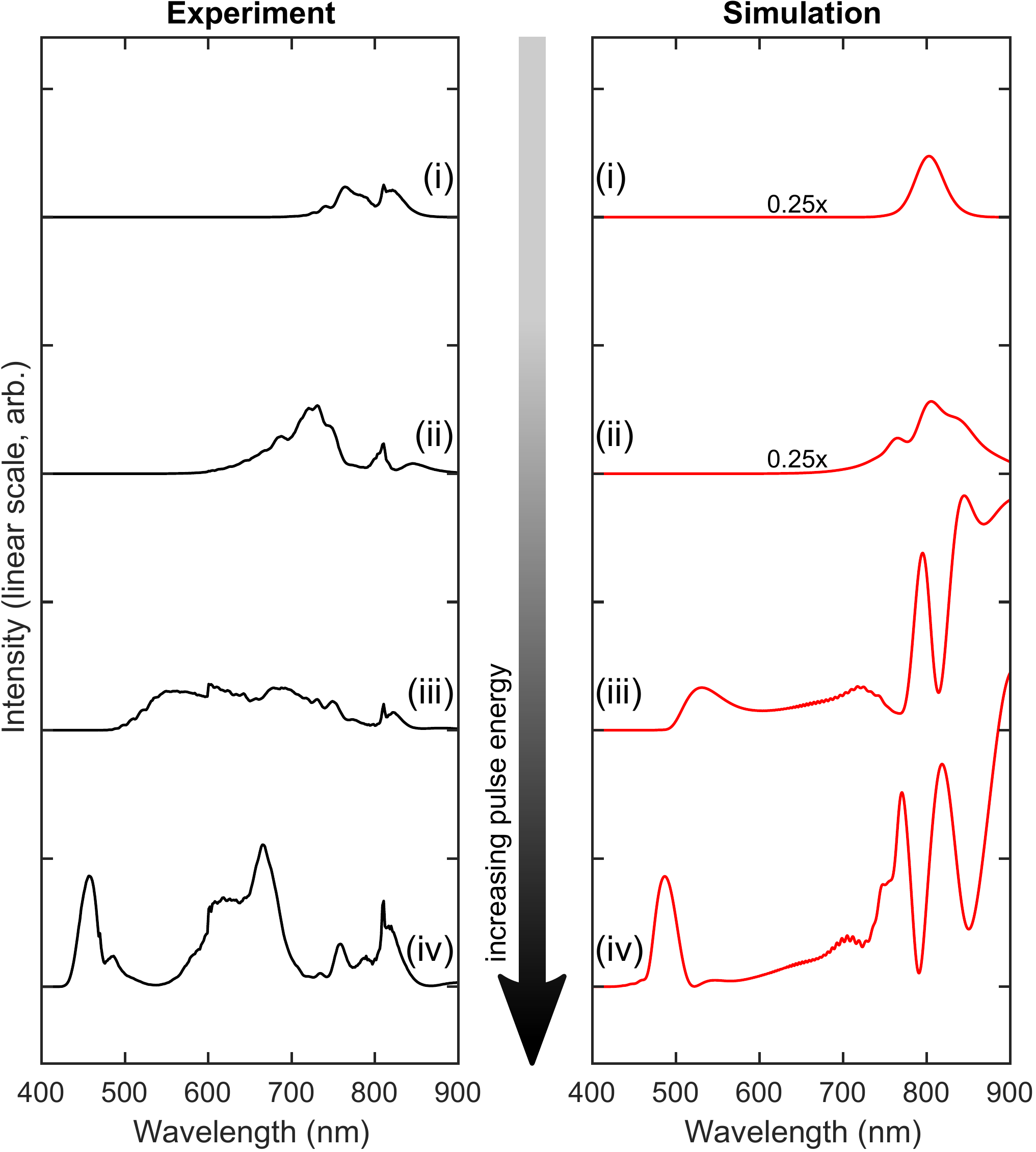}

\caption{\label{fig:expsim_comp}Comparison of experimental (left panel) and
simulated (right panel) output spectra from tapered PCF pumped by taccor source comb as a function of coupled pulse
energy. The spectra are offset for clarity. Except where noted, all
traces within each panel have the same scale and are normalized such
that the area under each curve is proportional to the pulse energy.
Experimental and simulated pulse energies for each row are, respectively,
(i)~24~pJ,~50~pJ, (ii)~63~pJ,~130~pJ, (iii)~100~pJ,~215
pJ, (iv)~190~pJ,~290~pJ. The discrepancies in the pulse energies
and spectral features are discussed in the text. }
\end{figure}

We compared our measured and simulated output spectra for the tapered PCF
(Fig.~\ref{fig:expsim_comp}) as a function of increasing pump power. The pulse energies shown in the right panel (simulations)
were chosen to produce representative spectra. In comparing simulations and measured spectra (Fig.~\ref{fig:expsim_comp}), 
we find two principal discrepancies: (1) the simulated pulse energies 
are a factor of $\sim\textrm{2}$ larger
than the corresponding experimental pulse energies, and (2) there
is very little radiation observed at wavelengths longer than the pump
wavelength in the experimental spectra. 
The first discrepancy may be due to the combined effect of (geometric) out-coupling loss and reduced infrared transmission through
the out-coupling objective, which is optimized for visible light; this mechanism lowers the measured out-coupled
power compared to simulated spectra since the power is measured after
the objective. These effects are thought to also contribute to the second discrepancy to some extent, but it is suspected that the dominant contribution comes from chromatic effects in coupling into the multimode fiber used with the OSA. Finally, there are uncertainties in fiber properties
(both geometric and optical) and laser parameters (e.g., coupled bandwidth)
which lead to uncertainties in the expected spectral profiles. 

Despite the differences
described above, the simulations qualitatively reproduce the trend in
the visible region quite well: initial symmetric broadening is 
followed by a wide, flat spectral plateau forming to the blue side
of the pump; also, at high pump powers, the blue-shifted radiation separates into
a distinct feature with a spectral gap between the pump
and the dispersive waves. Experimentally, $E_{p}\approx\textrm{100 pJ}$
is an ideal operating point for an astro-comb, because it produces
the flattest spectrum.

\begin{figure}[t]
\includegraphics[width=1\columnwidth]{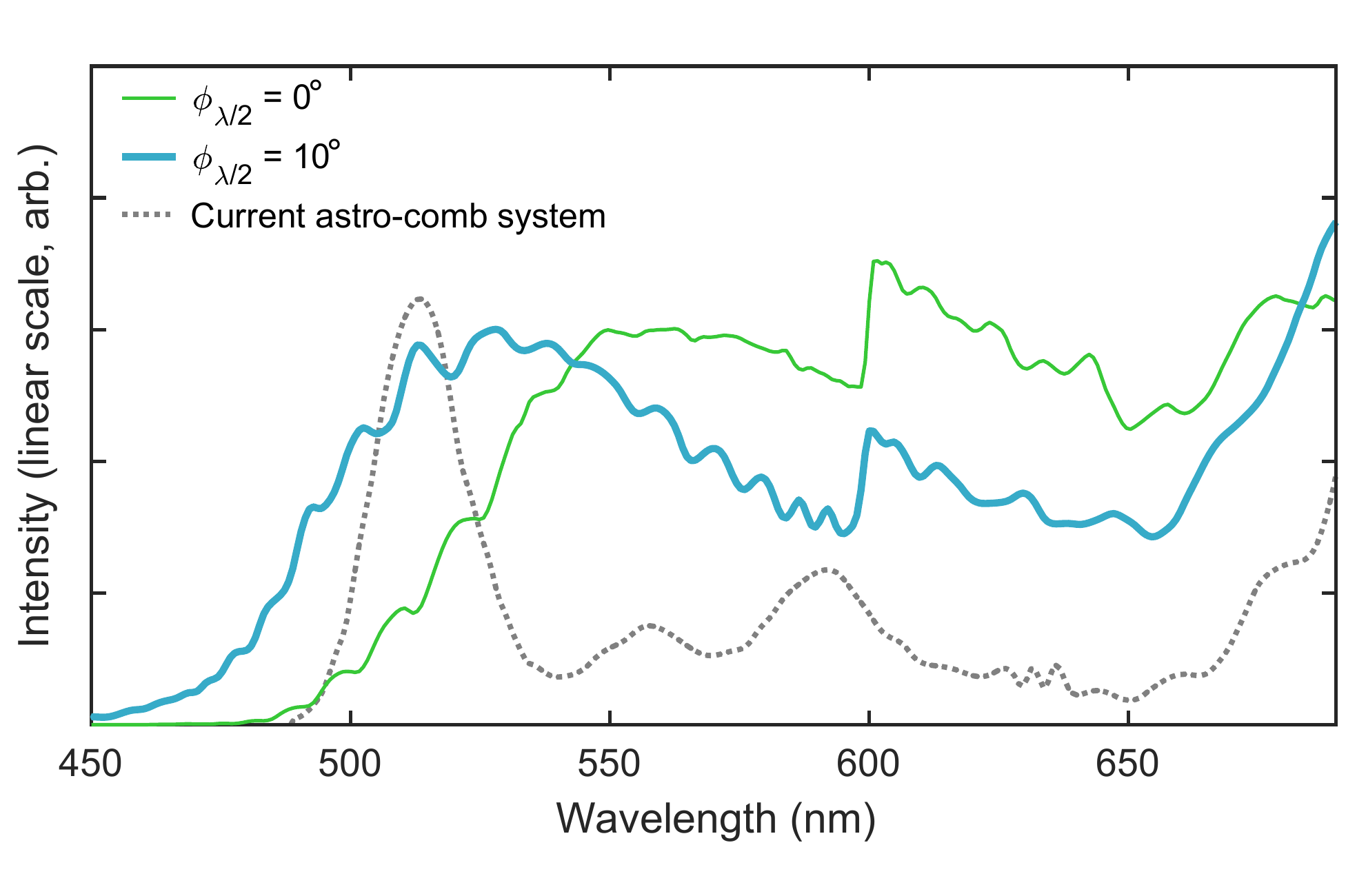}

\caption{\label{fig:poldep}Measured output spectra for tapered PCF pumped by taccor source comb at different input
polarizations (thin green and thick blue traces): a trade-off between spectral
bandwidth and flatness is observed by rotating the input polarization
using a half-wave plate at angle $\phi_{\lambda/2}$ (see text for
definition). The spectra were recorded with 100 pJ coupled into the tapered PCF.  The bandwidth and flatness of the green and blue traces
can be  compared to the PCF output spectrum from the currently
deployed astro-comb system \cite{Glenday2015} (dotted gray trace). }
\end{figure}

In a second test, we varied the input polarization using the half-wave plate
at constant coupled pump power. The results are
shown in Figure \ref{fig:poldep} (the kink in the thin green and thick blue
traces near 600 nm is thought to be an artifact from the OSA). The
condition denoted $\phi_{\lambda/2}=\textrm{0}^{\circ}$ indicates the wave plate angle
that gives the flattest spectrum at $E_{p}=\textrm{100 pJ}$ (this is
the angle at which all traces in the left panel of Figure
\ref{fig:expsim_comp} were recorded). At this nominal angle, the
intensity variation over the 530-690 nm range is only 1.2 dB. Rotating
the wave plate by 10 degrees results in a spectrum with increased
bandwidth at the expense of flatness: 3.2 dB variation over 490-690 nm,
with a blue tail extending below 450 nm. The polarization dependence likely comes from the fact that single-mode fibers support two modes with orthogonal polarizations. Natural birefringence \cite{Dudley2006} then leads to different output supercontinua for the two modes. This behavior cannot be quantitatively understood using our simulations as we use a model formulated using a single mode. A complete multi-mode formulation of the GNLSE \cite{Poletti2008} would
allow us to obtain more insight into the nature of these dynamics,
but this is beyond the scope of the current work.

\section{Conclusion}

Comparing the usable (visible) comb light from our new design to
the previous-generation astro-comb, which was based on a different
laser frequency comb and tapered PCF \cite{Glenday2015},
our new design provides  improved bandwidth and spectral flatness. We also estimate the power per comb mode by measuring the transmission through a 10 nm-wide bandpass filter around 532 nm, obtaining values of $\sim \textrm{10}^{\textrm{\footnotesize 2}}$~nW/mode, which is comparable to the result in Ref. \cite{Glenday2015}. It thus
satisfies the requirements for the astro-comb application.
Moreover, the new tapered PCF enables the use of a turn-key
Ti:sapphire laser, which greatly simplifies astro-comb design and operation \cite{Ravi2017}.

Our new astro-comb (employing the turn-key laser and tapered PCF described in this work) is expected to reach a radial velocity precision of \mbox{< 10 cm/s} (i.e., \mbox{< 200 kHz} in units of optical frequency) in a single exposure, required for detection of terrestrial exoplanets in the habitable zone around Sun-like stars. This is similar to results demonstrated in Ref. \cite{Ravi2017} using the same laser. Once the new astro-comb is fully deployed, we will verify the stability by injecting the astro-comb light into both channels of the astrophysical spectrograph simultaneously and compute the two-sample deviation of the spectral shift between channels as a function of averaging time \cite{Glenday2015,Ravi2017}. The next major step is to improve the residual dispersion of the Fabry-Perot mode filters \cite{Chen2010} used for repetition rate multiplication, so as to preserve all of the bandwidth generated by the PCF. Another viable option would be to split the comb light into several bands and filter each band separately using narrowband cavities \cite{Braje2008}.

In addition to the
calibration of astrophysical spectrographs used for Doppler velocimetry
of stars, our tapered PCF design may find applications in optical
coherence tomography (OCT) \cite{Hartl2001}. In OCT systems, the
axial (spatial) resolution scales as $\sim\lambda_{0}^{2}/\left(\Delta\lambda\right)$,
where $\lambda_{0}$ is the center wavelength and $\Delta\lambda$
is the bandwidth of the source. Hence, the resolution benefits from reducing
the center wavelength and increasing the bandwidth, which is a similar
design problem to the one addressed here. Ultrahigh-resolution visible-wavelength OCT has enabled optical sectioning at the subcellular level
\cite{Marchand2017} as well high-speed inspection of printed circuit
boards \cite{Czajkowski2012}. In such applications, spectral gaps
in the output band of supercontinuum sources used in OCT studies degrade
the axial resolution below that possible with the full band. Thus,
the spectral uniformity possible from the present tapered PCF design may improve resolution
further; it may also obviate some of the challenges associated with dual-band
OCT, where sophisticated signal-processing techniques are required
to combine information from spectrally separated bands~\cite{Cimalla2012}.

In summary, we demonstrated 
a tapered PCF that produces spectrally flat
light almost spanning the entire visible range when pumped by a turn-key
GHz Ti:sapphire laser. Our result represents a marked improvement in the
amount of optical bandwidth available to calibrate a visible-wavelength
spectrograph. This work also enables a simple 
visible frequency comb system without the need for spectral shaping.

\appendices
\begin{table*}[t]
\caption{Definitions for GNLSE model}\label{tab:def}
\centering
\begin{tabular}{p{0\linewidth}p{0.1\linewidth}p{0.5\linewidth}}
\hline 
\textbf{Symbol} &  & \textbf{Definition}\tabularnewline
\hline 
$\ast$ & \enskip{} & convolution\tabularnewline
$\mathcal{F}$$\left\{ f\left(T\right)\right\} $ &  & $=\int_{-\infty}^{\infty}f\left(T\right)\mathrm{e}^{\mathrm{i}\left(\omega-\omega_{0}\right)T}\textrm{d}T,$
Fourier operator\tabularnewline
$\mathcal{F}$$^{-1}\left\{ g\left(\omega-\omega_{0}\right)\right\} $ &  & $=\left(2\pi\right)^{-1}\int_{-\infty}^{\infty}g\left(\omega-\omega_{0}\right)\mathrm{e}^{-\mathrm{i}\left(\omega-\omega_{0}\right)T}\textrm{d}\omega,$
inverse Fourier operator\tabularnewline
$z$ &  & longitudinal coordinate\tabularnewline
$\omega$ &  & angular frequency\tabularnewline
$t$ &  & time\tabularnewline
$\omega_{\mathrm{\tiny 0}}$ &  & reference frequency (set to center of computational window)\tabularnewline
$\omega_{c}$ &  & pulse carrier frequency\tabularnewline
$\ensuremath{\beta\left(\omega\right)}$ &  & $=n_{\mathrm{eff}}\left(\omega\right)\times\omega/c,$ propagation constant\tabularnewline
$n_{\mathrm{eff}}\left(\omega\right)$ &  & effective index\tabularnewline
$c$ &  & speed of light in vacuum\tabularnewline
$\beta_{n}\left(\omega\right)$ &  & $=\partial^{n}\beta/\partial\omega^{n},$ $n^{\textrm{\tiny th}}$-order dispersion\tabularnewline
$T$ &  & $=t-\beta_{1}\left(\omega_{c}\right)z,$ time in comoving frame\tabularnewline
$\tilde{A}\left(z,\omega-\omega_{0}\right)$ &  & spectral envelope of pulse\tabularnewline
$A\left(z,T\right)$ &  & $=\mathcal{F}^{-1}\left\{ \tilde{A}\left(z,\omega-\omega_{0}\right)\right\} ,$
time-domain envelope of pulse\tabularnewline
$\mathcal{D}\left(\omega\right)$ &  & $=\mathrm{i}\left[\beta\left(\omega\right)-\beta\left(\omega_{0}\right)-\beta_{1}\left(\omega_{c}\right)\left(\omega-\omega_{0}\right)\right]-\frac{1}{2}\alpha\left(\omega\right),$
dispersion operator\tabularnewline
$\alpha\left(\omega\right)$ &  & frequency-dependent loss\tabularnewline
$\tilde{A}_{I}\left(z,\omega-\omega_{0}\right)$ &  & $=\textrm{e}^{-\mathcal{D\left(\omega\right)}z}\tilde{A}\left(z,\omega-\omega_{0}\right),$
interaction picture spectral envelope of pulse\tabularnewline
$\ensuremath{\bar{\gamma}\left(\omega\right)}$ &  & $=\frac{n_{2}n_{\mathrm{eff}}\left(\omega_{0}\right)\omega}{cn_{\mathrm{eff}}\left(\omega\right)A_{\mathrm{eff}}^{1/4}\left(\omega\right)}$,
frequency-dependent nonlinear parameter\tabularnewline
$n_{2}$ &  & nonlinear refractive index\tabularnewline
$A_{\mathrm{eff}}\left(\omega\right)$ &  & $=\frac{\left(\int_{-\infty}^{\infty}\int_{-\infty}^{\infty}\left|F\left(x,y,\omega\right)\right|^{2}dxdy\right)^{2}}{\int_{-\infty}^{\infty}\int_{-\infty}^{\infty}\left|F\left(x,y,\omega\right)\right|^{4}dxdy},$
frequency-dependent mode effective area\tabularnewline
$\ensuremath{F\left(x,y,\omega\right)}$ &  & transverse modal distribution\tabularnewline
$\bar{A}\left(z,T\right)$ &  & $=\mathcal{F}^{-1}\left\{ \tilde{A}\left(z,\omega-\omega_{0}\right)/A_{\mathrm{eff}}^{1/4}\left(\omega\right)\right\} $\tabularnewline
$R\left(t\right)$ &  & $=\left(1-f_{r}\right)\delta\left(t\right)+f_{r}h_{r}\left(t\right)\Theta\left(t\right),$
Raman response function\tabularnewline
$f_{r}$ &  & fractional contribution of delayed Raman response\tabularnewline
$h_{r}\left(t\right)$ &  & $=\frac{\tau_{1}^{2}+\tau_{2}^{2}}{\tau_{1}\tau_{2}^{2}}\exp\left(-t/\tau_{2}\right)\sin\left(t/\tau_{1}\right),$ with Raman fit parameters $\tau_{1},\tau_{2}$\tabularnewline
$\Theta\left(t\right)$ &  & Heaviside function\tabularnewline
\hline
\end{tabular}
\end{table*}
\section{Theoretical Model}
We describe the propagation of optical pulses in PCFs using the generalized
nonlinear Schr\"{o}dinger equation (GNLSE). Here, we work in the interaction picture and use a frequency domain  formulation, following Ref.~\cite{Hult2007}. The GNLSE \cite{Dudley2010} is expressed as
\setlength{\arraycolsep}{0.0em}
\begin{eqnarray}
\frac{\partial}{\partial z}\tilde{A}_{I}&{}={}&\mathrm{i}\bar{\gamma}\left(\omega\right)\exp\left(-\mathcal{D}\left(\omega\right)z\right)\nonumber\\
&&{\times}\:\mathcal{F}\left\{ \bar{A}\left(z,T\right)\times\left(R\left(T\right)\ast\left|\bar{A}\left(z,T\right)\right|^{2}\right)\right\}.\label{eq:propagation_eqn}
\end{eqnarray}
\setlength{\arraycolsep}{5pt}
Table \ref{tab:def} summarizes the definitions of all symbols used above.

In terms of the GNLSE, the PCF is entirely described 
by $\mathcal{D}\left(\omega\right)$,
$\bar{\gamma}\left(\omega\right)$ and $R\left(t\right)$.
The fiber material is fused silica, so we take
$f_{r}=\textrm{0.18}$, $\tau_{1}=\textrm{12.2 fs}$, and $\tau_{2}=\textrm{32 fs}$ as given in Ref. \cite{Agrawal2012}, and $n_2\approx\textrm{2.5}\times\textrm{10}^{-\textrm{\footnotesize20}}\textrm{ m}^{\textrm{\footnotesize2}}/\textrm{W}$ \bibnote{Crystal Fibre A/S, \href{https://www.thorlabs.com/\_sd.cfm?fileName=12663-M01.pdf\&partNumber=NL-2.8-850-02}{NL-2.8-850-02 datasheet}.}
for all our calculations. We neglect any losses, i.e., $\alpha\left(\omega\right)=0$ dB/m.

Note that in tapered geometries, both the
dispersion operator and the frequency-dependent nonlinear parameter
become functions of $z$ as well, i.e., $\mathcal{D}\left(\omega\right)\rightarrow\mathcal{D}\left(\omega,z\right)$
and $\bar{\gamma}\left(\omega\right)\rightarrow{\bar{\gamma}}\left(\omega,z\right)$
\cite{Travers2009,Judge2009}. This approach has been pointed out
to not be strictly correct as it does not conserve the photon number
\cite{Vanvincq2011,Laegsgaard2012}, but we have adopted it here for
simplicity. Our solver codes and data are available online \bibnote{\href{http://walsworth.physics.harvard.edu/code/scgen-taper.zip}{http://walsworth.physics.harvard.edu/code/scgen-taper.zip}}.

\section{Calculation of Supercontinuum Coherence}
We use a first-order measure $g_{12}^{\left(1\right)}$ to evaluate
the coherence \cite{Dudley2006} of the output optical field $\tilde{A}(\omega)$ from the PCF, 
\begin{equation}
\left|g_{12}^{\left(1\right)}\left(\omega\right)\right|=\frac{\left|\left\langle \tilde{A}_{i}^{\ast}\left(\omega\right)\tilde{A}_{j}\left(\omega\right)\right\rangle _{i\neq j}\right|}{\sqrt{\left\langle \left|\tilde{A}_{i}\left(\omega\right)\right|^{2}\right\rangle \left\langle \left|\tilde{A}_{j}\left(\omega\right)\right|^{2}\right\rangle }}
\end{equation}
where $\langle\cdots\rangle$ is an ensemble average over $N$ propagations
of the simulation.

Each run of the simulation differs by some
noise injected into the input field. We include only a shot
noise seed and no spontaneous Raman noise, as shot noise
has been shown to be the dominant noise process \cite{Corwin2003}.
To perturb the input pulse (in the time domain), we
follow Ref.~\cite{Paschotta2004,Ruehl2011}: to each temporal
bin of both the real and imaginary components of $A\left(T\right)$,
we add a random number drawn from a normal distribution with zero mean and
variance $\hbar\omega_c/\left(4\Delta T\right)$, where $\omega_c$ is
the pulse carrier frequency and $\Delta T$ is the temporal bin width.
In Figure \ref{fig:mainsim}c, we evaluate $\left|g_{12}^{\left(1\right)}\right|$
over $N=\textrm{100}$ runs of the simulation.

\section*{Acknowledgments}
The authors would like thank Guoquing Chang and Franz K\"{a}rtner
for supplying us with the NL-2.8-850-02 photonic crystal fiber for the project, as well as for their thoughtful reading of the manuscript. A.~Ravi
would also like to thank Pawel Latawiec, Fiorenzo Omenetto, Liane
Bernstein, Jennifer Schloss, Matthew Turner, Timothy Milbourne, G\'{a}bor
F\H{u}r\'{e}sz, and Tim Hellickson for helpful discussions.

This research work was supported by the Harvard Origins of Life Initiative,
the Smithsonian Astrophysical Observatory, NASA award no. NNX16AD42G,
NSF award no. AST-1405606. A.~Ravi was supported by a postgraduate scholarship from the Natural Sciences and Engineering Research Council of Canada.

\bibliographystyle{ieeetr}
\bibliography{references}

\vfill
\noindent\mbox{\footnotesize Authors' biography not available at the time of publication.}

\end{document}